\documentclass[]{aa}  

\usepackage{graphicx}
\usepackage{txfonts}
\usepackage{ragged2e}
\usepackage[colorlinks = true,
            linkcolor = blue,
            urlcolor  = blue,
            citecolor = blue,
            anchorcolor = blue]{hyperref}
\usepackage{url}
\usepackage[encapsulated]{CJK}

\begin{document} 

   \title{Investigating photometric and spectroscopic variability in the multiply-imaged `Little Red Dot' A2744-QSO1}

   \author{Lukas J. Furtak\inst{1}\thanks{\email{furtak@post.bgu.ac.il}},
           Amy R. Secunda\inst{2},
           Jenny E. Greene\inst{3},
           Adi Zitrin\inst{1},
           Ivo Labb\'{e}\inst{4},
           Miriam Golubchik\inst{1},
           Rachel Bezanson\inst{5},
           Vasily Kokorev\inst{6},
           Hakim Atek\inst{7},
           Gabriel B. Brammer\inst{8,9},
           Iryna Chemerynska\inst{7},
           Sam E. Cutler\inst{10},
           Pratika Dayal\inst{11},
           Robert Feldmann\inst{12},
           Seiji Fujimoto\inst{6},
           Karl Glazebrook\inst{4},
           Joel Leja\inst{13,14,15},
           Yilun Ma (\begin{CJK*}{UTF8}{gbsn}马逸伦\end{CJK*})\inst{3},
           Jorryt Matthee\inst{16},
           Rohan P. Naidu\inst{17},
           Erica J. Nelson\inst{18},
           Pascal A. Oesch\inst{19,8},
           Richard Pan\inst{20},
           Sedona H. Price\inst{5},
           Katherine A. Suess\inst{18},
           Bingjie Wang (\begin{CJK*}{UTF8}{gbsn}王冰洁\end{CJK*})\inst{13,14,15},
           John R. Weaver\inst{10}
           \and
           Katherine E. Whitaker\inst{10,8}
          }

   \institute{Department of Physics, Ben-Gurion University of the Negev, P.O. Box 653, Be'er-Sheva 84105, Israel
   \and
   Center for Computational Astrophysics, Flatiron Institute, New York, NY 10010, USA
   \and
   Department of Astrophysical Sciences, Princeton University, Princeton, NJ 08544, USA
   \and
   Centre for Astrophysics and Supercomputing, Swinburne University of Technology, Melbourne, VIC 3122, Australia
   \and
   Department of Physics \& Astronomy and PITT PACC, University of Pittsburgh, Pittsburgh, PA 15260, USA
   \and
   Department of Astronomy, The University of Texas at Austin, Austin, TX 78712, USA
   \and
   Institut d'Astrophysique de Paris, CNRS, Sorbonne Universit\'e, 98bis Boulevard Arago, 75014, Paris, France
   \and
   Cosmic Dawn Center (DAWN), Copenhagen, Denmark
   \and
   Niels Bohr Institute, University of Copenhagen, Jagtvej 128, Copenhagen, Denmark
   \and
   Department of Astronomy, University of Massachusetts, Amherst, MA 01003, USA
   \and
   Kapteyn Astronomical Institute, University of Groningen, PO Box 800, 9700 AV Groningen, The Netherlands
   \and
   Department of Astrophysics, Universit\"{a}t Z\"{u}rich, Winterthurerstrasse 190, CH-8044 Zurich, Switzerland
   \and
   Department of Astronomy \& Astrophysics, The Pennsylvania State University, University Park, PA 16802, USA
   \and
   Institute for Computational \& Data Sciences, The Pennsylvania State University, University Park, PA 16802, USA
   \and
   Institute for Gravitation and the Cosmos, The Pennsylvania State University, University Park, PA 16802, USA
   \and
   Institute of Science and Technology Austria (ISTA), Am Campus 1, 3400 Klosterneuburg, Austria
   \and
   MIT Kavli Institute for Astrophysics and Space Research, 70 Vassar Street, Cambridge, 02139, Massachusetts, USA
   \and
   Department for Astrophysical and Planetary Science, University of Colorado, Boulder, CO 80309, USA
   \and
   D\'{e}partement d’Astronomie, Universit\'{e} de Gen\`{e}ve, Chemin Pegasi 51, 1290 Versoix, Switzerland
   \and
   Department of Physics and Astronomy, Tufts University, 574 Boston Ave., Medford, MA 02155, USA
   } 

  \date{Received 17 February 2025; accepted 2 May 2025}
 
  \abstract{JWST observations have uncovered a new population of red, compact objects at high redshifts dubbed `Little Red Dots' (LRDs), which typically show broad emission lines and are thought to be dusty Active Galactic Nuclei (AGN). Some of their other features, however, challenge the AGN explanation, such as prominent Balmer breaks and extremely faint or even missing metal high-ionization lines, X-ray, or radio emission, including in deep stacks. Time variability is another, robust, test of AGN activity. Here, we exploit the $z=7.045$ multiply-imaged LRD A2744-QSO1, which offers a particularly unique test of variability due to lensing-induced time delays between the three images spanning 22\,yr (2.7\,yr in the rest-frame), to investigate its photometric and spectroscopic variability. We find the equivalent widths (EWs) of the broad H$\alpha$ and H$\beta$ lines, which are independent of magnification and other systematics, to exhibit significant variations, up to $18\pm3$\,\% for H$\alpha$ and up to $22\pm8$\,\% in H$\beta$, on a timescale of 875\,d (2.4\,yr) in the rest-frame. This suggests that A2744-QSO1 is indeed an AGN. We find no significant photometric variability beyond the limiting systematic uncertainties, so it currently cannot be determined whether the EW variations are due to line-flux or continuum variability. These results are consistent with a typical damped random walk (DRW) variability model for an AGN like A2744-QSO1 ($M_{\mathrm{BH}}=4\times10^7\,\mathrm{M}_{\odot}$) given the sparse sampling of the light-curve with the available data. Our results therefore support the AGN interpretation of this LRD, and highlight the need for further photometric and spectroscopic monitoring in order to build a detailed and reliable light-curve.}


   \titlerunning{Variability of the multiply-imaged LRD A2744-QSO1}
   \authorrunning{L. J. Furtak et al.}

   \maketitle
%

\section{Introduction} \label{sec:intro}
One of the most prominent results from the JWST \citep{gardner23,mcelwain23} in its first few years of operations is the serendipitous discovery of a new population of extremely red point-sources at high redshifts ($z\gtrsim4$), the so-called `Little Red Dots' \citep[LRDs; e.g.][]{matthee24a}. These were initially identified in JWST imaging data \citep[e.g.][]{endsley23,labbe23,labbe25,furtak23d,barro24} with follow-up spectroscopy later revealing broad emission lines, consistent with heavily dust-attenuated ($A_V\gtrsim3$), broad-line type~1 active galactic nuclei \citep[AGN; e.g.][]{kocevski23,harikane23b,matthee24a,furtak24b,kokorev23,greene24,killi24}, although the picture has grown more complicated with time \citep[e.g.][]{perez-gonzalez24,baggen24,baggen24E}. Many LRD AGN have been observed to date, down to $z\sim2.5$ \citep[e.g.][]{wang24a,kocevski24,naidu24}, with the largest photometric samples encompassing several hundred objects selected in JWST's blank fields \citep[e.g.][]{kokorev24a,kocevski24}. At $z\gtrsim4$, LRDs seem to represent 20\,\% to 30\,\% of the AGN population \citep[e.g.][]{kocevski24} and tend to reside in over-densities \citep[e.g.][]{schindler24,matthee24b}. If LRDs are indeed powered by an accreting black hole, they pose a challenge for black hole growth models as they present extreme black-hole-to-galaxy mass ratios $M_{\mathrm{BH}}/M_{\star}\gtrsim0.03$ \citep[e.g.][]{kokorev23,furtak24b,durodola24,matthee24b} which are not observed in the low-redshift Universe (though cf.\ \citealt{lin25,euclid25_q1-LRDs}).

Most LRDs are, however, extremely faint in X-rays \citep[e.g.][]{ananna24,yue24} or radio \citep[][]{mazzolari24,perger25,gloudemans25}, and seem to lack a hot dust component in the mid-infrared \citep[e.g.][]{williams24,akins24,setton25}. These features could be explained with e.g.\ Compton-thick broad-line regions \citep[][]{maiolino25}, super-Eddington accretion \citep[e.g.][]{pacucci24,lambrides24,inayoshi24,madau24,madau25,tripodi24}, or seeds of primordial origin \citep[e.g.][]{dayal24}. However, they have also, in combination with the detection of distinct Balmer-breaks in some LRDs \citep[e.g.][]{wang24b,setton24,ma25,labbe24}, led to alternative models where their spectral energy distribution (SED) is dominated by stars \citep[e.g.][]{perez-gonzalez24,wang24b}. While the extreme stellar densities implied in such a picture could a priori help explain the observed broad lines \citep{baggen24,baggen24E}, such extreme densities are not otherwise observed in the Universe. On the other hand, the presence of a Balmer-break does not necessarily contradict an AGN interpretation, as material around the black hole can in theory have physical conditions such that a significant fraction of the hydrogen atoms have their electrons in or above the second energy level \citep[e.g.][]{laor11,inayoshi25,li25,ji25,naidu25}.

Since all type~1 AGN vary temporally \citep[e.g.][]{matthews63,kelly09,macleod12,cammelli25}, variability could be a key discriminator between AGN and stellar-only models. Recent studies have therefore also looked for evidence of variability in LRDs \citep{kokubo24,zhang24,tee25,ji25}. The investigated samples show mostly no variability in several epochs of broad-band photometry, both in the rest-frame ultra-violet (UV) and optical \citep[][]{kokubo24,zhang24,tee25} with only a small subsample of LRDs displaying potential variability \citep{zhang24}. However, since JWST has only been operational for a little over two years, time-domain data are still sparse. Therefore, variability studies of LRDs so far are based on a handful of JWST epochs or rely on archival \textit{Hubble Space Telescope} (HST) data \citep[e.g.][]{tee25}, which only probe the faint rest-frame UV emission of LRDs.

In this work, we examine the variability of the multiply-imaged LRD A2744-QSO1 \citep{furtak23d,furtak24b}, one of the first LRDs ever detected. It was first identified photometrically as a uniquely red-and-compact object in the \textit{Ultra-deep NIRSpec and NIRCam ObserVations before the Epoch of Reionization} \citep[UNCOVER;][]{bezanson24} observations of the strong lensing (SL) cluster Abell~2744 \citep[A2744; e.g.][]{abell89,merten11} and then spectroscopically confirmed as a broad-line LRD at $z_{\mathrm{spec}}=7.0451\pm0.0001$ \citep{furtak23d,furtak24b}. Being multiply-imaged, A2744-QSO1 is ideal for this type of study because SL induces a delay in the relative arrival time of each of the three images. This means that each visit effectively provides three epochs of observation, separated by several years. With two years of JWST imaging and spectroscopy of A2744-QSO1 available to date, we can investigate it for variability in its emission over a two-decade timescale (see \citet{ji25} for another recent analysis of variability in this object; we also compare to their results in our discussion).

This letter is organized as follows: The data are described in section~\ref{sec:data}. We then present our spectroscopic and photometric analyses in sections~\ref{sec:spectroscopy} and~\ref{sec:photometry}, respectively, and then discuss our results and their implications in section~\ref{sec:discussion}. We assume a flat $\Lambda$CDM cosmology with $H_0=70$ km s$^{-1}$ Mpc$^{-1}$, $\Omega_{\Lambda}=0.7$, and $\Omega_\mathrm{m}=0.3$, and use AB magnitudes \citep{oke83}. Errors are typically $1\sigma$ unless stated otherwise.

\section{Data} \label{sec:data}

\begin{table}
\label{tab:images}
\caption{The three images of A2744-QSO1 and their gravitational magnifications $\mu$ and time delays $\Delta t$.}
\begin{tabular}{lccccccc}
\hline\hline
Image   &   R.A.~[Deg.]       & Dec.~[Deg.]     &   $\Delta t_{\mathrm{grav}}~[\mathrm{d}]$    &   $\mu$\\\hline
A       &   $3.5798408$       & $-30.4015681$   &   $0$                                        &   $5.4\pm1.1$\\
B       &   $3.5835462$       & $-30.3966783$   &   $855_{-6}^{+72}$                           &   $7.5\pm1.5$\\
C       &   $3.5972175$       & $-30.3943403$   &   $7043_{-152}^{+234}$                       &   $2.7\pm0.5$\\\hline
\hline
\end{tabular}
\footnotesize
\par\noindent -- \texttt{Note:} See also \citet{furtak23d}. The gravitational magnifications and time delays are computed from the \texttt{v2} UNCOVER SL model \citep{furtak23c,price25}.
\end{table}

\begin{table}
\label{tab:epochs}
\caption{List of epochs of JWST observations available for A2744-QSO1 to date.}
\begin{tabular}{lccc}
\hline\hline
Epoch   &   Program ID(s)  &   PI(s)                    &    Date [MJD]\\\hline
\multicolumn{4}{c}{\textit{JWST/NIRCam imaging}}\\\hline
1       &   2756           &   Chen                     &   59872.56\\
2       &   2561           &   Labb\'{e} \& Bezanson    &   59890.23\\
3       &   2756           &   Chen                     &   59919.42\\
4       &   2883, 3538     &   Sun, Iani                &   60247.37\\
5       &   4111           &   Suess                    &   60262.06\\
6$^{a}$ &   3516           &   Naidu \& Matthee         &   60283.75\\\hline
\multicolumn{4}{c}{\textit{JWST/NIRSpec spectroscopy}}\\\hline
7       &   2561           &   Labb\'{e} \& Bezanson    &   60157.53\\
8$^{b}$ &   2561           &   Labb\'{e} \& Bezanson    &   60522.16\\\hline\hline
\end{tabular}
\par\smallskip
\footnotesize
$^{a}$Single-filter epoch for F356W from ALT \citep{naidu24}.
\par $^{b}$Repeat observation, only for image~A.
\end{table}

\begin{figure*}
    \centering
    \includegraphics[width=\textwidth]{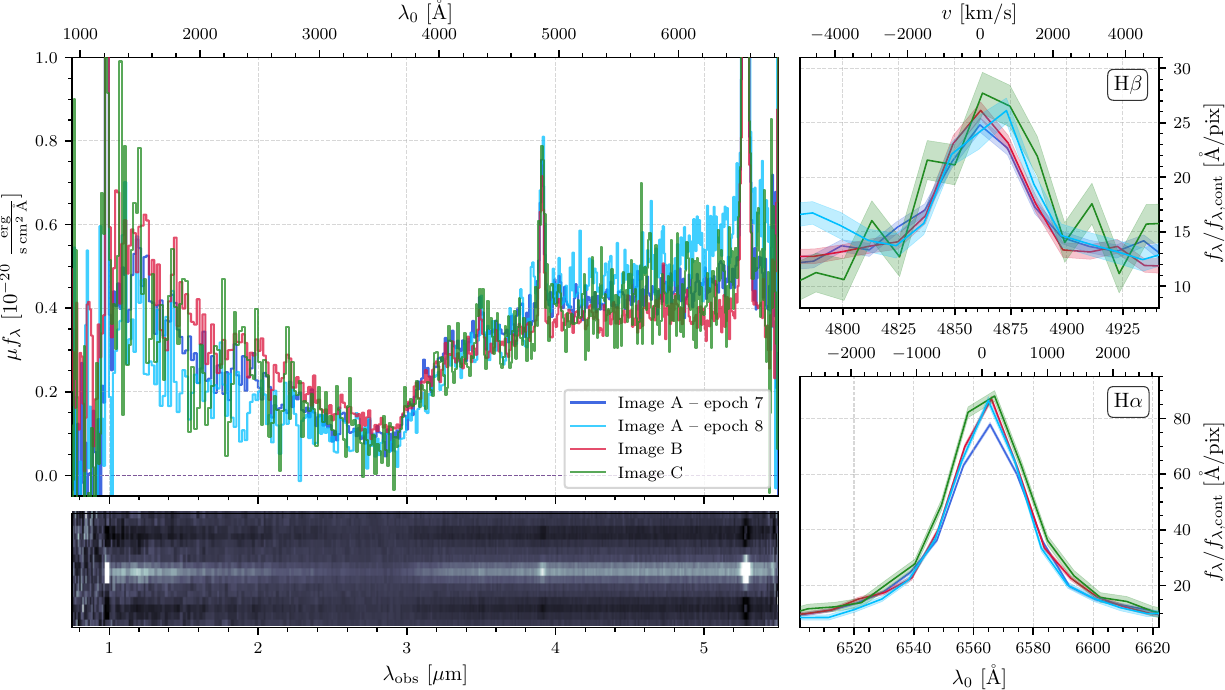}
    \caption{JWST/NIRSpec prism spectroscopy from UNCOVER of all three images of A2744-QSO1 in both spectroscopic epochs (see Tab.~\ref{tab:epochs}). Image~A is shown in blue (dark for epoch~7, light for epoch~8), image~B in red and image~C in green. \textit{Left}: Full NIRSpec-prism spectra, normalized to the total luminosity of image~A in epoch~7, obtained by integrating the spectra in wavelengths $1.0\,\mu\mathrm{m}<\lambda_{\mathrm{obs}}<5.4,\mu\mathrm{m}$. Thanks to the new reductions (see section~\ref{sec:data}), the broad H$\alpha$ line no longer falls off the detector as in previous work, but is seen entirely here for the first time. \textit{Right}: Zoom-ins on the H$\alpha$ (\textit{bottom}) and H$\beta$ (\textit{top}) lines, scaled by their continua listed in Tab.~\ref{tab:line-fluxes} such that they become independent of magnification, similar to the EW.}
    \label{fig:spectra}
\end{figure*}

In the two years since the first JWST observations of UNCOVER, the center of A2744 was imaged numerous times with JWST. Here, we use the \texttt{Grism redshift \& line analysis software for space-based slitless spectroscopy} \citep[\texttt{grizli};][]{grizli23} to query the \texttt{Mikulski Archive for Space Telescopes} (\texttt{MAST}) for \textit{Near Infrared Camera} \citep[NIRCam;][]{rieke23} imaging targeted at the cluster center, and to perform the data reduction. Each filter and epoch is drizzled into mosaics of 0.04\arcsec\ per pixel. We list our six imaging epochs in Tab.~\ref{tab:epochs}. The NIRCam observations reduced here include data from programs 2756 (PI: W.~Chen), 2561 \citep[UNCOVER; PIs: I.~Labb\'{e} \& R.~Bezanson;][]{bezanson24}, 2883 (\textit{Magnif}; PI: F.~Sun), 3538 (PI: E.~Iani), 4111 \citep[\textit{MegaScience}; PI: K.~Suess;][]{suess24}, and 3516 \citep[\textit{All the Little Things}; PIs: R.~Naidu \& J.~Matthee;][ALT]{naidu24}.

Our spectroscopic analysis is based on data from the UNCOVER program \citep{price25}, which also obtained ultra-deep \textit{Near Infrared Spectrograph} \citep[NIRSpec;][]{jakobsen22,boeker23} prism observations using the micro-shutter array \citep[MSA;][]{ferruit22,jakobsen24} in July~2023. Due to a guide-star acquisition failure in the initial observations, a second epoch of prism spectra was taken in July~2024, which includes image~A of A2744-QSO1 in the MSA. For this work, we utilize \texttt{v4} reductions from the \texttt{DAWN JWST Archive} \citep[\texttt{DJA};][]{heintz24,degraaff24b}, which are reduced using the development version (\texttt{v0.9.5}) of \texttt{MSAEXP} \citep{brammer23}. We refer the reader to \citet{price25} for most of the details of the \texttt{MSAEXP} reduction and refer to appendix~\ref{app:spectroscopy} for more information. \ The primary improvement of the new reductions is a substantial extension of the wavelength coverage to include the full NIRSpec prism disperser throughput to cover wavelengths $0.6\,\mu\mathrm{m}<\lambda<5.5\,\mu\mathrm{m}$, achieving spectral resolutions $R\sim30-300$. This has the advantage of fully covering the H$\alpha$ line for our target on the red end of the detector (see section~\ref{sec:spectroscopy}). While several JWST grism spectroscopy programs targeted A2744 as well, namely GLASS-JWST \citep[ID 1324;][]{treu22}, \textit{Magnif}, and ALT, GLASS-JWST does not cover wavelength ranges with emission lines for our target, and \textit{Magnif} only covers the H$\gamma$ line which has too low a signal-to-noise ratio (SNR) to be exploited here. The ALT spectroscopy covers the H$\beta$ line, but only shows marginal detections in two images and is therefore not usable for our variability studies.

Gravitational magnifications $\mu$ and time delays $\Delta t_{\mathrm{grav}}$, listed in Tab.~\ref{tab:images}, are computed analytically from the UNCOVER lens model \texttt{v2}, which was constructed by \citet{furtak23c} using a revised version of the \citet{zitrin15a} parametric method and was recently updated with JWST/NIRSpec spectroscopic redshifts in \citet{price25}. We assume a floor of 20\,\% uncertainty in the magnification to account for lensing systematics\footnote{In practice the systematics estimated for these magnification values can reach $\sim$40\,\% \citep{zitrin15a}. However, the exact level does not affect the conclusions of this work.} \citep[e.g.][]{zitrin15a}. While the time-delays are also prone to the same systematics, we find these to be of the order of a few hundred days at most, which does not affect the search for variability conducted in this work.

\section{Spectroscopic analysis} \label{sec:spectroscopy}

\begin{table}
    \label{tab:line-fluxes}
    \caption{JWST/NIRSpec emission line results for each image of A2744-QSO1 in each epoch of available spectroscopy (see Tab.~\ref{tab:epochs}).}
    \begin{center}
    \begin{tabular}{l|cccc}
    \hline\hline
    Line        &  \multicolumn{2}{c}{Image~A}                          &   Image~B                &   Image~C\\\hline
                &   Epoch~7                 &   Epoch~8                 &                          &   \\\hline
                &   \multicolumn{4}{c}{$\mathrm{EW}_0$~[\AA]}\\\hline
    H$\beta$    &   $48_{-2}^{+2}$          &   $53_{-4}^{+4}$          &   $50_{-2}^{+2}$         &   $62_{-5}^{+5}$\\
    H$\alpha$   &   $285_{-6}^{+7}$         &   $289_{-8}^{+9}$         &   $307_{-5}^{+4}$        &   $346_{-11}^{+11}$\\\hline
                &   \multicolumn{4}{c}{$F_{\mathrm{line}}~[10^{-19}\,\mathrm{erg}\,\mathrm{cm^{-2}\,s^{-1}}]$}\\\hline
    H$\beta$    &   $2.7\pm0.2$             &   $2.9\pm0.3$             &   $2.4\pm0.3$            &   $2.9\pm0.3$\\
    H$\alpha$   &   $20.2\pm0.8$            &   $24.7\pm1.1$            &   $18.3\pm1.4$           &   $24.7\pm1.1$\\\hline
                &   \multicolumn{4}{c}{$f_{\lambda,\mathrm{continuum}}~[10^{-22}\,\mathrm{erg}\,\mathrm{cm^{-2}\,s^{-1}\,\mathrm{\AA}^{-1}}]$}\\\hline
    H$\beta$    &   $6.9\pm0.3$             &   $6.7\pm0.3$             &   $5.6\pm0.2$            &   $4.8\pm0.4$\\
    H$\alpha$   &   $8.8\pm0.3$             &   $10.6\pm0.4$            &   $6.7\pm0.2$            &   $6.6\pm0.5$\\\hline\hline
    \end{tabular}
    \end{center}
    \footnotesize
    \par\noindent -- \texttt{Note:} The line fluxes and underlying continuum flux densities are de-magnified, but the EWs are independent of magnification.
\end{table}

\begin{figure}
    \centering
    \includegraphics[width=\columnwidth]{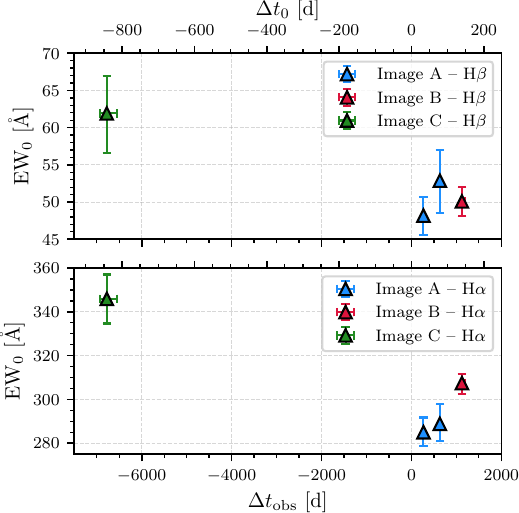}
    \caption{Spectroscopic variability of A2744-QSO1 as a function of time, with image~A shown in blue, image~B in red, and image~C in green as in Fig.~\ref{fig:multi-image_variability}. --\textit{Top}: H$\beta$ rest-frame EW, presenting a drop of $22\pm8$\,\% between images~C and~A. --\textit{Bottom}: H$\alpha$ rest-frame EW, presenting a drop of $18\pm3$\,\% in the time between images~C and~A. We note that the variability is consistent in both broad Balmer lines, as can be expected since they both originate from the broad-line region of A2744-QSO1. The EWs represent excellent indicators of variability since they are insensitive to most systematics such as lensing, slit-losses, calibration, etc.}
    \label{fig:EW-variability}
\end{figure}

With the UNCOVER JWST/NIRSpec spectra of the three images, and the repeat observation spectrum of image~A (see Tab.~\ref{tab:epochs}), we effectively have four epochs of spectroscopy for A2744-QSO1, spanning rest-frame wavelengths from Lyman-$\alpha$ (Ly$\alpha$) to H$\alpha$. The spectrum presents no other permitted (i.e.\ broad) emission lines other than the hydrogen Balmer series and Ly$\alpha$ \citep[see e.g.][]{furtak24b}. All four spectra are shown in Fig.~\ref{fig:spectra}, with a special focus on H$\alpha$ and H$\beta$, which we investigate for variability because they have the most robust SNRs. Thanks to the new spectroscopic reductions (see section~\ref{sec:data}), H$\alpha$ no longer falls off the detector as in our previous work on this object \citep[][see also \citealt{ji25}]{furtak24b}, which enables us to probe the variability at a much higher SNR than with H$\beta$ alone.

We measure the line fluxes by fitting spectra as explained in detail in appendix~\ref{app:spectroscopy}. The resulting integrated emission line fluxes, underlying continua, and rest-frame equivalent-widths (EWs) are listed in Tab.~\ref{tab:line-fluxes} for each image and epoch. Unlike absolute line fluxes, the EWs are neither prone to lensing nor instrumental systematics (e.g.\ slit-losses), which is why we use them as the primary and most robust indicators of variability \citep[see also][]{ji25}. As can be seen in Fig.~\ref{fig:EW-variability}, both H$\alpha$ and H$\beta$ show a consistent variability in their EW over time\footnote{Here and throughout this letter, we define the time axis in days relative to the UNCOVER imaging (epoch~2 at $\mathrm{MJD}=59890.23$, see Tab.~\ref{tab:epochs}). The relative time is then obtained by adding the difference between the epochs and $\Delta t_{\mathrm{grav}}$, the gravitational time delay listed in Tab.~\ref{tab:images}.}. More quantitatively, we note a drop in EW of $18\pm3$\,\% in H$\alpha$ and $22\pm8$\,\% in H$\beta$ between the times probed by image~C and image~A (in both its epochs). The absolute calibration of the red edge of the NIRSpec prism, where H$\alpha$ is located, is uncertain ($5-10$\,\%), but this should not impact the EW measurements directly. Finally, we do not find any variability in the broad line widths within the uncertainties.

Four data points do not, however, sufficiently sample the light-curve to robustly characterize the spectroscopic variability of this source or to determine if the variations in the broad-line EWs arise from line-flux variations, continuum flux variations, or both. If driven by continuum variability, these EW variations would translate to $|\Delta m|\sim0.2$\,magnitudes, i.e.\ far below the uncertainty floor in our photometric analysis over the same timescale (section~\ref{sec:photometry}), and thus would not be detected. Our results nonetheless clearly show significant variability in A2744-QSO1, in particular given the high SNR in the H$\alpha$ line, and confirm the results recently found by \citet{ji25} using H$\beta$ alone. We take this as an indication that the broad lines in A2744-QSO1 indeed originate from an AGN. It is hard to imagine a systematic effect that would simultaneously affect the EWs of both lines.

\section{Photometric analysis} \label{sec:photometry}
We also investigate A2744-QSO1 for photometric variability. In section~\ref{sec:photometric-lightcurve} we present its photometric light-curve, and then assess what level of variability can be expected with dedicated simulations in section~\ref{sec:simulations}.

\subsection{Observed photometric light-curve} \label{sec:photometric-lightcurve}

\begin{figure*}
    \centering
    \includegraphics[width=\textwidth]{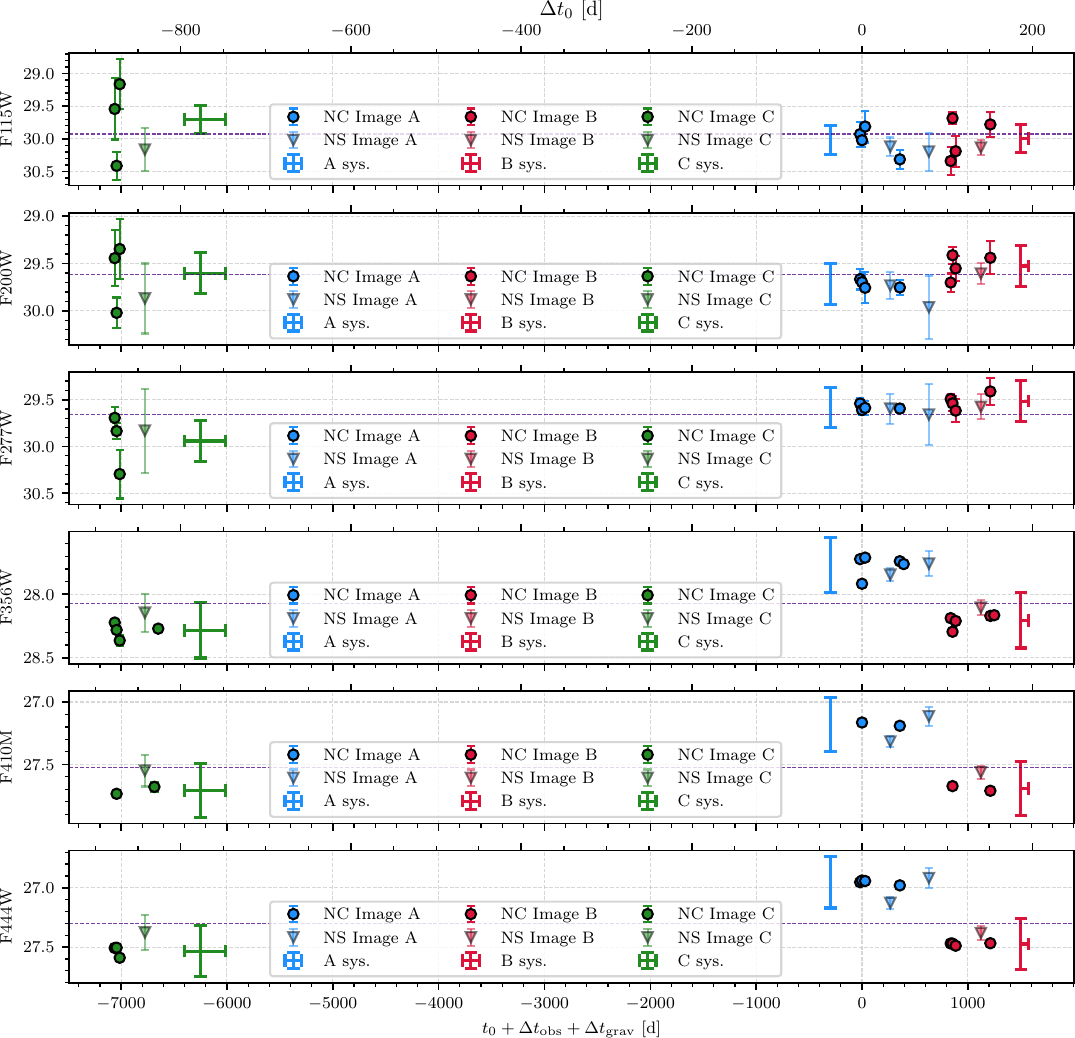}
    \caption{De-magnified light-curve of A2744-QSO1 in each photometric band with at least one epoch in both 2022 and 2023, as a function of time, incorporating the gravitational time delays $\Delta t_{\mathrm{grav}}$ given in Tab.~\ref{tab:images}. Image~A is shown in blue, Image~B in red, and Image~C in green. Solid circles represent NIRCam (NC) data-points and solid triangles represent the synthetic photometry obtained by integrating the NIRSpec (NS) spectra in the broad-band filter band-passes. The dashed horizontal line represents the mean $\bar{m}$ from which variations $|\Delta m|$ are calculated in each band. Note that the magnification uncertainties are not propagated into the photometry here since for each image, the magnification remains constant  and is only relevant when comparing to the other images. Instead, the colored crosses indicate the level of SL (magnification and time-delay) systematics for each image. We also note that image~B lies close to a cluster galaxy which might contaminate its photometry.}
    \label{fig:multi-image_variability}
\end{figure*}

Since we are dealing with three perfect point-sources in all NIRCam bands \citep[see][]{furtak23d}, we measure simple aperture photometry of the three images in all bands and epochs in a consistent way and list the fluxes in Tab.~\ref{tab:flux}. The details on the flux measurements can be found in appendix~\ref{app:photometry}. In addition to the NIRCam photometry, we integrate the NIRSpec spectra in the NIRCam filter band-passes to gain four additional effective epochs of photometry in each band, also listed in Tab.~\ref{tab:flux}.

In Fig.~\ref{fig:multi-image_variability} we show the photometric light-curve of A2744-QSO1 in selected filters chosen to have at least one epoch in each of the November windows of 2022 and 2023, thus bridging the one-year gap. The F115W-, F200W-, and F277W-bands, which cover the rest-frame UV emission of our object, present small NIRCam flux variations which are fully consistent with the photometric systematics (see Fig.~\ref{fig:RMS}). Likewise, we see no variability in rest-frame optical photometry for the three images taken alone (F356W-, F410M-, and F444W; Fig.~\ref{fig:RMS}). We do find variations up to $|\Delta m|\sim0.4$\,magnitudes\footnote{Note, the $|\Delta m|=\mathrm{max}(m-\bar{m})$ is defined here as the maximum difference to the mean across our entire light-curve in each band (dashed horizontal lines in Fig.~\ref{fig:multi-image_variability}).} in the full light-curve covering longer timescales (22\,yr, i.e.\ 2.7\,yr in the rest-frame).

While the individual error-bars on each point of photometry are very small, in particular in the optical bands, our measurement is dominated by systematics. The main source of uncertainty in this case are the magnification errors, of 0.22\,magnitudes (20\,\%) chosen to represent the typical systematic uncertainties associated with SL mass modeling (see section~\ref{sec:data}). Our measurement is furthermore prone to significant additional systematics of $\sim0.05-0.3$\,magnitudes at the observed brightnesses of the images (see Fig.~\ref{fig:RMS}), arising from varying depths, coverage and observing strategy between the epochs. We also note that image~B lies close to a cluster galaxy (see \citealt{furtak23d}), meaning that its photometry and background estimate might be contaminated.

Thus, the variations that we see in the photometry of A2744-QSO1 are at best marginal detections. The current data do not yield a robust measurement of photometric variability beyond the uncertainties, at least not with the current sampling of the light-curve. Nonetheless, given the limited temporal sampling and the stochastic nature of AGN variability, this does not necessarily imply there to be no intrinsic variability in the source, as discussed in section~\ref{sec:simulations}.

\subsection{Simulated light-curves} \label{sec:simulations}

\begin{figure}
    \centering
    \includegraphics[width=\columnwidth]{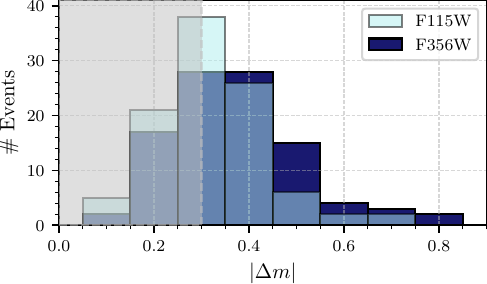}
    \caption{Histogram of maximum variation $|\Delta m|$ from 100 simulated DRW light-curves of an $M_{\mathrm{BH}}=4\times10^7\,\mathrm{M}_{\odot}$ black hole, sampled at the rest-frame effective times of our combined observations of all three images of A2744-QSO1 as shown in Fig.~\ref{fig:multi-image_variability}. Rest-frame UV (F115W) variations are shown in light blue and optical variations (F356W) are shown in dark blue. We choose to show the F356W-band here because it has the most epochs (see Tab.~\ref{tab:flux}) and therefore has the highest probability of detecting a significant event. The gray-shaded area represents variations that are below the typical errors of 0.3\,magnitudes (including systematics) in our photometry measurements in both bands. Even with the sampling of our full light-curve, the probability to detect a significant variation event ($|\Delta m|>0.5$\,magnitudes) is relatively low.}
    \label{fig:mock-lightcurve_prediction}
\end{figure}

\begin{figure*}
    \centering
    \includegraphics[width=\textwidth]{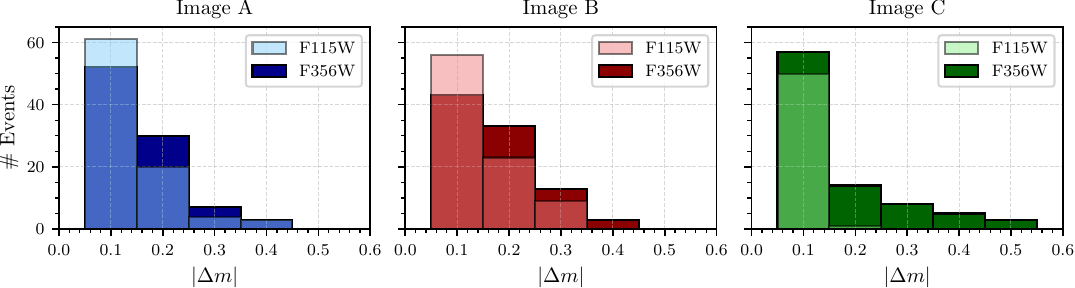}
    \caption{Histogram of maximum variation $|\Delta m|$ from our simulated light-curves (see Fig.~\ref{fig:mock-lightcurve_prediction}) in F115W (light) and F356W (dark), now sampled at the rest-frame observation times of each image individually. As in Fig.~\ref{fig:mock-lightcurve_prediction}, we show the F356W-band to represent the rest-frame optical because it has the most epochs. Image~A is shown in blue, image~B in red and image~C in green. The probabilities of observing a significant event are much lower for the individual images than when combining the images to a full light-curve.}
    \label{fig:mock-lightcurve_single-images}
\end{figure*}

In order to assess what can be expected in terms of broad-band variability from an AGN like A2744-QSO1, we simulate 100 light-curves using a damped random walk (DRW) model as in \citet{kelly09}. The DRW has been shown to be a decent representation of quasar light-curves for timescales of days to years \citep[e.g.][]{kozlowski10,macleod10,zu13}. It is a stochastic process that characterizes the variability power-spectrum as a power-law with exponent $-2$ at high frequencies and white noise at frequencies below a certain damping timescale $\tau_{\mathrm{DRW}}$. We use $\tau_{\mathrm{DRW}}=76$\,d, which corresponds to the damping timescale for an AGN with mass $M_{\mathrm{BH}}=4\times10^7\,\mathrm{M}_{\odot}$ \citep[corresponding to A2744-QSO1;][]{furtak24b} as parameterized by \citet{burke21}, and a structure function at infinity of $\mathrm{SF}_{\infty}=0.3$, which is a typical value for \textit{Sloan Digital Sky Survey} (SDSS) quasars \citep{macleod10}. We then sample each light-curve at the relative (rest-frame) observation times of each epoch and image for the F115W-band in the UV, and the three optical bands F356W, F410M, and F444W, because they have the highest SNRs. We add Gaussian noise to the mock observations scaled by the SNRs of the observations.

The resulting distribution of $|\Delta m|$ per mock light-curve is shown in Fig.~\ref{fig:mock-lightcurve_prediction} for all three images combined. While we find the probability for observing a $|\Delta m|>0.3$\,magnitude event, i.e., slightly above the average uncertainties, to be of 56\,\%, 67\,\%, 53\,\%, and 56\,\% in the F115W-, F356W-, F410M-, and F444W-bands, respectively, the probability to observe a significant event, e.g.\ $|\Delta m|>0.5$\,magnitudes, is of only 6\,\% (F115W), 17\,\% (F356W), 15\,\% (F410M), and 6\,\% (F444W).

We also break-up our mock light-curves into the three separate images, which have two to five epochs each, in Fig.~\ref{fig:mock-lightcurve_single-images}. The probability of a significant event in the F356W-band, i.e.\ $|\Delta m|>0.3$\,magnitudes given the dispersion between epochs at the observed brightness of each image shown in Fig.~\ref{fig:RMS}, is $<13$\,\% and $<3$\,\% for a $|\Delta m|>0.5$\,magnitude event. In all other bands, the probability is $<6$\,\% for a $|\Delta m|>0.3$\,magnitude event, and a $|\Delta m|>0.5$\,magnitude event cannot be observed for any of the mock light curves in these filters. In our highest SNR band, F444W, a significant event in the individual images would be $|\Delta m|>0.2$\,magnitude. Given our sampling of the light-curve, the probabilities to detect such an event there would however also only be of $1-17$\,\%.

The difference in the likelihood of detecting variability in the various wavebands depends only on the SNR of the observations and the specific observation times in each band. We find the likelihood of detecting variability is greatest for the F356W-band because it has the largest number of observations. However, if we take the wavelength dependence of $\mathrm{SF}_{\infty}$ from \citet{burke23} into account, the probability of detecting variability in the F356W-band and the other optical wavebands decreases significantly. Finally, if we hold $\mathrm{SF}_{\infty}$ constant, for $M_{\mathrm{BH}} \leq 10^6\,\mathrm{M}_{\odot}$, the probability of a significant excursion of $|\Delta m|=0.2-0.3$\,magnitudes for the F356W- or F444W-bands becomes $\sim70-100$\,\%, making such low masses seem less likely for A2744-QSO1.

\section{Discussion and conclusion} \label{sec:discussion}
Our variability measurements in the multiply-imaged LRD A2744-QSO1 come after several years of debate as to the nature of LRDs, which started when they were first identified photometrically \citep[e.g.][]{labbe23,labbe25,furtak23d,barro24}, and has continued as more and more of the puzzling aspects of their SEDs were revealed \citep[e.g.][]{perez-gonzalez24,ananna24,yue24,setton24,gloudemans25}. Because of the elusive nature of the LRDs' SED \citep[e.g.][]{setton24,labbe24}, at this stage we cannot predict which part of the SED -- the optical or the UV continuum -- can be expected to vary in the first place. The hardest component of the SED to explain with a purely stellar model are the broad emission lines -- both their high luminosity and their width. And indeed, we do find these to vary consistently in H$\alpha$ and H$\beta$, by $18\pm3$\,\% and $22\pm8$\,\%, respectively as shown in section~\ref{sec:spectroscopy}. Note that \citet{ji25} find a similar result for H$\beta$ in the same source, which we here confirm via the higher SNR H$\alpha$ line. The current data, however, are insufficient to securely determine whether the EW variations are dominated by variations in the continuum, the line fluxes, or perhaps most probably -- both. If dominated by continuum variation, our detection of broad-line EW variability in section~\ref{sec:spectroscopy} would correspond to $|\Delta m|\sim0.2$\,magnitude excursions, which are below the limiting sensitivity of our photometric analysis in section~\ref{sec:photometric-lightcurve}.

While the last year in particular saw several studies looking into photometric variability of LRDs, the vast majority of LRDs did not display significant variations over the handful of epochs considered \citep[][]{kokubo24,zhang24,tee25}, similar to what we find for each image individually. Only eight out of several hundred investigated LRDs seem to present significant variability \citep{zhang24}. This could a priori indicate the LRD population at large to not necessarily be AGN, although it has also been claimed that e.g.\ super-Eddington accretion could explain the lack of variability \citep[e.g.][]{inayoshi24}. In the context of these non-detections in the literature, our photometric results do not come as a surprise. Even when combining the three images, and thus tripling the number of effective epochs to $11-14$, we do not find any significant photometric variability, i.e.\ large $|\Delta m|>0.5$\,magnitude excursions (see section~\ref{sec:photometric-lightcurve}). This is nonetheless still consistent with the stochastic nature of AGN variability \citep[e.g.][]{kelly09,macleod10}, as we have shown in section~\ref{sec:simulations}. Indeed, our DRW simulations show that even if there were intrinsic photometric variability in A2744-QSO1, the probability of detecting it with our sampling is low and nearly impossible in the small amount of epochs for each image separately. The lack of significant photometric variability in LRDs in the literature therefore cannot rule out intrinsic variability of LRDs, nor that they are AGN. It merely means that with the data available so far, the light-curves of LRDs are not sufficiently sampled to rule out expected levels of photometric variability.

Our detection of significant broad-line EW variability in the multiply-imaged $z=7$ LRD A2744-QSO1 presents strong evidence in support of the AGN interpretation of this object, and a varying AGN at that, even if we do not find significant photometric variability at this stage. At the time of writing, high-resolution JWST/NIRSpec spectroscopy of A2744-QSO1 has revealed strong narrow absorption features in the Balmer-lines \citep[][]{ji25,deugenio25}, which further corroborate the scenario where this object is dominated by an AGN, though of lower black-hole mass than previous estimates, with a nearly absent host galaxy \citep[$M_{\mathrm{BH}}/M_{\mathrm{dyn}}>0.02-0.4$;][]{deugenio25}. Most recently, the detection of two peculiar LRD objects, with extreme Balmer-breaks and a complete absence of metal lines as in A2477-QSO1 \citep[][]{naidu25,degraaff25}, has further shed light on the nature of LRDs: The extreme Balmer-break, broad lines, and absorption features in said lines are consistent with emission from a super-massive black hole embedded in an extremely dense cocoon of hot gas, similar to a stellar atmosphere \citep[e.g.][]{inayoshi25,ji25,naidu25}, which in addition suppresses the radio and X-ray emission \citep[e.g.][]{lambrides24,inayoshi24,maiolino25,rusakov25}. As demonstrated in detail in \citet{naidu25}, the combination of such a `black hole star' (BH*) with a (very weak) host galaxy, to make up the rest-frame UV emission, reproduces the observed spectrum of A2744-QSO1 quite well. This scenario also perfectly explains our simultaneous detection of broad-line EW variability, which come from the BH*, and non-detection of rest-frame UV variability, which would originate from young stars forming in the vicinity of the BH*.

Our results further show that robust investigations into the variability of LRDs, including A2744-QSO1, will require dedicated monitoring with numbers of epochs large enough to sample the light-curve frequently and maximize the probability to detect significant variation events. More importantly, we stress that such monitoring campaigns need to be consistent in depth, coverage, and observing strategy to minimize the systematics between single epochs. Thanks to the gravitational time delays, this source is ideally suited for further study of its variability and could make for a prime target for a large-scale reverberation mapping campaign in the future. With that, one could potentially measure the black-hole mass of an AGN at $z=7$ directly and, for the first time, calibrate the scaling relations used to measure black-hole masses beyond the near-by Universe \citep[e.g.][]{liu24}. Thanks to SL time delays, multiply-imaged, cluster-lensed AGN like A2744-QSO1 may open a door to achieve this on a reasonable timescale out to high-redshifts and high black-hole masses (e.g.\ \citealt{golubchik24}).

\begin{acknowledgements}
      We would like to thank Xihan Ji, Hannah \"{U}bler, and Roberto Maiolino, for cordial and useful discussions. The BGU lensing group acknowledges support by grant No.~2020750 from the United States-Israel Binational Science Foundation (BSF) and grant No.~2109066 from the United States National Science Foundation (NSF), and by the Israel Science Foundation Grant No.~864/23. P.D.\ warmly thanks the European Commission's and University of Groningen's CO-FUND Rosalind Franklin program.

      This work is based on observations obtained with the NASA/ESA/CSA JWST, namely programs GO-2756, -2561, -2883, -3538, -4111, and -3516, retrieved from the \texttt{Mikulski Archive for Space Telescopes} (\texttt{MAST}) at the \textit{Space Telescope Science Institute} (STScI). STScI is operated by the Association of Universities for Research in Astronomy, Inc. under NASA contract NAS 5-26555. The spectroscopy products presented herein, from JWST program GO-2561, were retrieved from the \textit{Dawn JWST Archive} (\texttt{DJA}). \texttt{DJA} is an initiative of the \textit{Cosmic Dawn Center} (DAWN), which is funded by the Danish National Research Foundation under grant DNRF140. The data used in this work may be retrieved from the \texttt{MAST} archive at: \url{http://dx.doi.org/10.17909/p7t7-te67}. This work also makes use of the Center for Computational Astrophysics at the Flatiron Institute which is supported by the Simons Foundation. Support for JWST programs GO-2561, -4111, and -3516 was provided by NASA through grants from STScI.

      This research made use of \texttt{Astropy},\footnote{\url{http://www.astropy.org}} a community-developed core Python package for Astronomy \citep{astropy13,astropy18} and \texttt{Photutils}, an \texttt{Astropy} package for detection and photometry of astronomical sources \citep{photutils_v1.13.0}, as well as the packages \texttt{NumPy} \citep{vanderwalt11}, \texttt{SciPy} \citep{virtanen20}, \texttt{Matplotlib} \citep{hunter07}, and the \texttt{MAAT} Astronomy and Astrophysics tools for \texttt{MATLAB} \citep[][]{maat14}.
\end{acknowledgements}

\bibliographystyle{aa} 
\bibliography{references} 

\begin{appendix}
\onecolumn
\section{Photometric measurements and systematic uncertainties} \label{app:photometry}

\begin{table*}[b!]
    \label{tab:flux}
    \caption{De-magnified photometry of A2744-QSO1 used in section~\ref{sec:photometry} to investigate the source's variability.}
    \begin{center}
    \begin{tabular}{l|cccccc}
    \hline\hline
    Image   &   F115W               &   F200W               &   F277W               &   F356W               &   F410M               &   F444W\\\hline
            &   \multicolumn{6}{c}{\textit{NIRCam epoch~1} -- GO-2756}\\\hline
    A       &   $29.927\pm0.191$    &   $29.666\pm0.108$    &   $29.541\pm0.058$    &   $27.722\pm0.010$    &   -                   &   $26.952\pm0.007$\\
    B       &   $30.338\pm0.210$    &   $29.702\pm0.095$    &   $29.492\pm0.054$    &   $28.188\pm0.013$    &   -                   &   $27.470\pm0.009$\\
    C       &   $29.542\pm0.467$    &   $29.444\pm0.294$    &   $29.693\pm0.119$    &   $28.223\pm0.031$    &   -                   &   $27.570\pm0.022$\\\hline
            &   \multicolumn{6}{c}{\textit{NIRCam epoch~2} -- GO-2561}\\\hline
    A       &   $30.021\pm0.101$    &   $29.697\pm0.084$    &   $29.611\pm0.052$    &   $27.917\pm0.012$    &   $27.164\pm0.009$    &   $26.939\pm0.006$\\
    B       &   $29.684\pm0.088$    &   $29.414\pm0.089$    &   $29.539\pm0.076$    &   $28.945\pm0.018$    &   $27.675\pm0.010$    &   $27.471\pm0.008$\\
    C       &   $30.412\pm0.211$    &   $30.021\pm0.161$    &   $29.835\pm0.081$    &   $28.281\pm0.019$    &   $27.736\pm0.018$    &   $27.505\pm0.012$\\\hline
            &   \multicolumn{6}{c}{\textit{NIRCam epoch~3} -- GO-2756}\\\hline
    A       &   $29.812\pm0.243$    &   $29.757\pm0.162$    &   $29.587\pm0.073$    &   $27.711\pm0.012$    &   -                   &   $26.943\pm0.008$\\
    B       &   $30.189\pm0.237$    &   $29.553\pm0.135$    &   $29.617\pm0.121$    &   $28.210\pm0.029$    &   -                   &   $27.488\pm0.014$\\
    C       &   $29.163\pm0.381$    &   $19.378\pm0.320$    &   $30.295\pm0.261$    &   $28.364\pm0.041$    &   -                   &   $27.589\pm0.028$\\\hline
            &   \multicolumn{6}{c}{\textit{NIRCam epoch~4} -- GO-2883 \& GO-3538}\\\hline
    A       &   $30.313\pm0.145$    &   $29.753\pm0.080$    &   $29.595\pm0.051$    &   $27.739\pm0.009$    &   $27.191\pm0.014$    &   $26.980\pm0.005$\\
    B       &   $29.779\pm0.194$    &   $29.440\pm0.176$    &   $29.411\pm0.141$    &   $28.170\pm0.033$    &   $27.712\pm0.023$    &   $27.468\pm0.018$\\
    C       &   -                   &   -                   &   -                   &   -                   &   $27.681\pm0.041$    &   -\\\hline
            &   \multicolumn{6}{c}{\textit{NIRCam epoch~6} -- GO-3516}\\\hline
    A       &   -                   &   -                   &   -                   &   $27.761\pm0.012$    &   -                   &   -\\
    B       &   -                   &   -                   &   -                   &   $28.165\pm0.014$    &   -                   &   -\\
    C       &   -                   &   -                   &   -                   &   $28.271\pm0.026$    &   -                   &   -\\\hline\hline
            &   \multicolumn{6}{c}{\textit{NIRSpec synthetic epoch~7} -- GO-2561}\\\hline
    A       &   $29.947\pm0.146$    &   $29.631\pm0.146$    &   $29.477\pm0.158$    &   $27.752\pm0.051$    &   $27.265\pm0.045$    &   $27.101\pm0.052$\\
    B       &   $29.863\pm0.120$    &   $29.417\pm0.110$    &   $29.405\pm0.135$    &   $27.978\pm0.059$    &   $27.450\pm0.053$    &   $27.341\pm0.061$\\
    C       &   $30.165\pm0.421$    &   $29.788\pm0.398$    &   $29.850\pm0.504$    &   $28.119\pm0.156$    &   $27.604\pm0.135$    &   $27.435\pm0.155$\\\hline
            &   \multicolumn{6}{c}{\textit{NIRSpec synthetic epoch~8} -- GO-2561}\\\hline
    A       &   $30.329\pm0.348$    &   $30.009\pm0.345$    &   $29.727\pm0.334$    &   $27.854\pm0.099$    &   $27.232\pm0.081$    &   $27.025\pm0.090$\\
    B       &   -                   &   -                   &   -                   &   -                   &   -                   &   -               \\
    C       &   -                   &   -                   &   -                   &   -                   &   -                   &   -               \\\hline
    \hline
    \end{tabular}
    \end{center}
    \footnotesize
    \par\noindent -- \texttt{Note:} A summary of epochs is given in Tab.~\ref{tab:epochs}. Note that the uncertainties here do not contain systematics such as magnification uncertainties or the systematic scatter between epochs. The filters listed here are only those used in section~\ref{sec:photometry}, i.e.\ the bands that have coverage in at least one epoch in 2022 and 2023 each. The NIRCam epoch~5, listed in Tab.~\ref{tab:epochs}, only contains medium bands that are not observed in any other epoch and is therefore not used further in section~\ref{sec:photometry}.
\end{table*}

\begin{figure*}
    \centering
    \includegraphics[width=\textwidth]{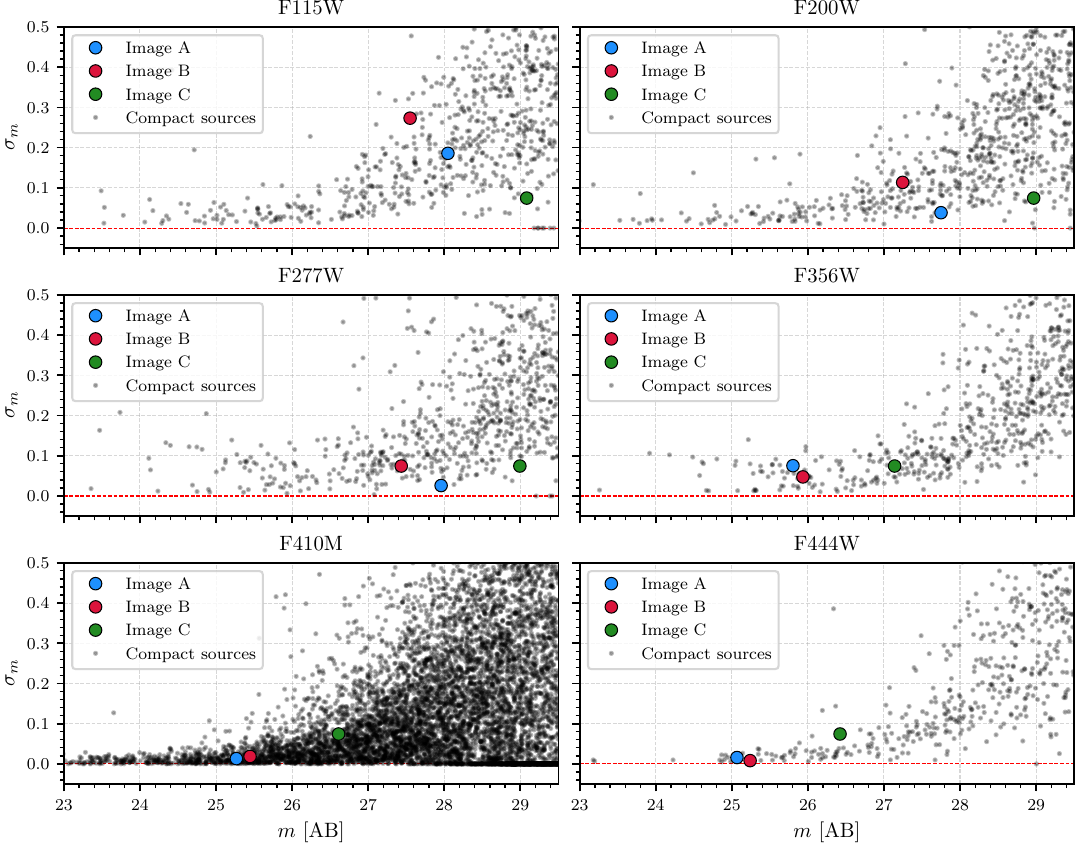}
    \caption{Systematic dispersion (standard deviation) $\sigma_m$ across all epochs of NIRCam imaging as a function of observed magnitude for each band used in our analysis in section~\ref{sec:photometry}. The black dots represent compact ($z<0.4\arcsec$) sources selected from the \citet{weaver24} UNCOVER catalog, chosen to be covered by each epoch. Image~A is shown in blue, image~B in red, and image~C in green. A2744-QSO1 does not present photometric variability beyond the typical dispersion of objects in the catalog objects across all epochs.}
    \label{fig:RMS}
\end{figure*}

In order to have consistent photometry measurements across all three images and all epochs, we measure photometry in fixed circular apertures of 0.16\arcsec\ radius with \texttt{photutils} \citep[\texttt{v1.13.0}][]{photutils_v1.13.0} in all epochs and bands, without re-centering the aperture between measurements. The latter ensures a more robust sampling of the error-space. The background is measured as the sigma-clipped median in circular annulii of inner and outer radii 0.4\arcsec\ and 0.7\arcsec\ and is subtracted from the flux measurement. Both the measurement aperture and the background annulus are carefully chosen to avoid any possible contamination in all three images. Note that since all three images are point-sources in all NIRCam bands \citep[see][]{furtak23d}, our measurements are agnostic to differential magnification. In Tab.~\ref{tab:flux}, we list the fluxes of each image in each filter and epoch used in section~\ref{sec:photometry}. While F070W and F090W observations are also available from e.g.\ \textit{MegaScience} \citep[][]{suess24} and ALT \citep{naidu24}, the object is not detected in these bands due to its Lyman-break and we therefore do not consider them further here. The fluxes are de-magnified with the nominal magnification factors listed in Tab.~\ref{tab:images}, but without propagating the magnification uncertainties and other systematics. We however estimate our measurements to be dominated by significant systematics from lensing and between the multiple epochs as discussed in section~\ref{sec:photometry} and in the following. We also integrate the NIRSpec spectra in the NIRCam band-passes for extra epochs of photometry as mentioned in section~\ref{sec:photometry}. The resulting flux values are also listed in Tab.~\ref{tab:flux}.

Since our epochs of observation comprise JWST/NIRCam imaging data from several different programs (see Tab.~\ref{tab:epochs} and section~\ref{sec:data}), which were all targeted at different science cases and therefore have vastly differing coverages in area and filters, depths, and observing strategies, our photometric light-curves are prone to significant systematics, in addition to the magnification uncertainties, that are not taken into account in the flux uncertainties listed in Tab.~\ref{tab:flux}. In an attempt to quantify these systematics, we use the public UNCOVER photometric catalog\footnote{DR3, including the additional broad and medium bands from the \textit{MegaScience} program \citep{suess24}: \url{https://jwst-uncover.github.io/\#releases}.} by \citet{weaver24} to select compact sources ($a<0.4\arcsec$) in the A2744-field with magnitudes $20<m<29.5$. This selection is further re-fined with the \texttt{use\_phot} and \texttt{flag\_nearbcg} flags \citep[see][]{weaver24}. With a catalog of reference objects in hand, we cross-match their positions with the footprints of the data per band and epoch and keep only objects that are observed in every epoch in each band. We then compute the dispersion (standard deviation) in magnitude $\sigma_m$ across all epochs in each filter for each object as a quantification of the systematics between epochs. These are shown as a function of magnitude in Fig.~\ref{fig:RMS} for the six filters used in our analysis in section~\ref{sec:photometry}, alongside the same dispersion measurement for the three images of A2744-QSO1. As can be seen in the figure, the systematics between epochs are larger than any expected photometric variations in the three rest-frame UV bands according to the simulations that we conducted in section~\ref{sec:discussion}. In the optical bands, the systematics are somewhat lower for our source because it has much higher SNRs in those filters, but still of similar order as typically expected variations. None of the individual images show any variation above the systematic floor at its observed magnitude, which is consistent with what could be expected for DRW AGN variability (again see section~\ref{sec:discussion}). This shows that a robust investigation of the photometric variability of A2744-QSO1, and LRDs in general, requires a dedicated monitoring program with consistent depths and observing strategies to minimize these systematics.

\section{Spectroscopic measurements} \label{app:spectroscopy}
For spectroscopy, we use the latest reductions of the UNCOVER spectroscopic data \citep{price25} available from the \texttt{DJA} \citep{heintz24,degraaff24b}. The spectra are reduced with \texttt{MSAEXP} \texttt{v0.9.5} and reach $5.5\,\mu$m, fully covering H$\alpha$ at $z=7.04$. The spectra of images~A and~B have typical $\mathrm{SNR}=15-20$ per pixel in the continuum, while the spectrum of images~C, and~A in epoch~8, have $\mathrm{SNR}=8-10$. In addition to the uncertainties provided by \texttt{MSAEXP}, we adopt a systematic error floor of 3\,\% per pixel in the continuum and 6\,\% per pixel at the location of strong lines. These errors are added in quadrature before fitting the spectra to account for spectro-photometric calibration uncertainties and other uncertainties related to the detailed shape of the intrinsic lines and the NIRSpec line-spread-function (LSF). The spectra are modeled over an observed wavelength-range $\lambda=3.1-5.4\,\mu$m, i.e.\ red-ward of the red-shifted Balmer break, as we are here primarily interested in a robust measurement of the strongest Balmer lines and the continuum at the location of said lines.

The model fitted to the spectra consists of the following components: The continuum is modeled with a power-law slope $\beta$, convolved with an exponential dust law with index $-0.7$ and $0<A_V<6$. This simple model reproduces the curvature of the continuum well over the considered wavelength range, better than a power-law of variable slope. Note that fitting with the latter only changes the results of H$\alpha$ and H$\beta$ by $<1.5$\,\%. The H$\beta$ line is modeled with a single broad-line Gaussian profile since the spectral resolution at the observed wavelength ($R\sim320$) is not sufficient to resolve the narrow component, as was already shown in \citet{furtak24b}. H$\alpha$ on the other hand has both a higher SNR and a higher spectral resolution. It is therefore more complicated, requiring a broad-line and a narrow-line component, even at the JWST/NIRSpec prism resolution (effectively $R>500$ at $5\,\mu$m for point sources, see \citealt{degraaff24a}). Fainter lines adjacent to H$\beta$ (the [O\,\textsc{III}]$\lambda\lambda4959,5007$\AA-doublet, He\,\textsc{II}$\lambda4686$\AA, and H$\gamma$) are masked. The [N\,\textsc{II}]$\lambda\lambda6548,6583$\AA-doublet is undetectable and including it in the fit changes the H$\alpha$ by $<$1\%, which is why we omit it. The model is convolved with the wavelength-dependent NIRSpec prism LSF for a point-source \citep[][]{degraaff24a}. Finally, we include two nuisance parameters: One is a rescaling of the resolution of the spectrum by a factor $lsf\_scale=1.0-2.0$ with a uniform prior, reflecting uncertainty in the JWST/NIRSpec LSF. A value of $1.0$ is appropriate for uniformly exposed shutters and $1.8-2.0$ may be appropriate for point sources \citep{degraaff24a}. The second nuisance parameter is $noise\_scale=0.1-10$ with a logarithmic prior, reflecting a possible over- or underestimate of the noise in the fit. We list all free parameters and their priors in Tab.~\ref{tab:fit-priors}. The posterior distributions are then sampled with {\tt PyMultiNest} \citep{buchner14}.

First, we fit the spectrum of image~A (epoch~7), which has the highest SNR, with the broad- and narrow-line widths as free parameters, finding $\mathrm{FWHM}_{\mathrm{broad}}=2533\pm120$\,km/s for both broad lines, $\mathrm{FWHM}_{\mathrm{narrow}}=800\pm60$\,km/s, and $lsf\_scale=1.51$. In the subsequent fitting of the spectra of images~B, C, and~A (epoch~8), we fix the line widths and LSF scaling to the values of image~A to ensure consistency. Typical values of noise\_scale are $noise\_scale=1.0-1.15$, indicating that the errors in the spectrum were only slightly underestimated. Finally, from the best-fit model posterior samples, we determine for H$\alpha$ and H$\beta$ the continuum at the line center, the total line flux, and the EW, taking into account parameter correlations. For H$\alpha$, we sum the narrow and broad components as they are difficult to separate reliably. Basic convergence tests were done (live points, sampling efficiency), showing the measurements were robust. The resulting continuum fluxes, integrated line fluxes and EWs are listed in Tab.~\ref{tab:line-fluxes} in section~\ref{sec:spectroscopy}.

\begin{table*}
\label{tab:fit-priors}
    \caption{Free parameters of our spectral fit with their priors.}
    \begin{center}
    \begin{tabular}{lcc}
    \hline\hline
    Parameter                                                   &   Prior type  &   Prior bounds\\\hline
    $z$                                                         &   uniform     &   $[6.99,7.09]$\\
    $A_V$                                                       &   uniform     &   $[0,6]$\\
    $\beta$                                                     &   uniform     &   $[-2.4,5.0]$\\
    $L_{\mathrm{continum}}~[\mathrm{L}_{\odot}]$                &   log-uniform &   $[10^5,10^{12}]$\\
    $\mathrm{FWHM}_{\mathrm{broad}}$~[km/s]                     &   uniform     &   $[1000,5000]$\\
    $\mathrm{FWHM}_{\mathrm{narrow}}$~[km/s]                    &   uniform     &   $[100,900]$\\
    $L_{\mathrm{H}\alpha,\mathrm{broad}}~[\mathrm{L}_{\odot}]$  &   log-uniform &   $[10^5,10^{12}]$\\
    $L_{\mathrm{H}\beta,\mathrm{broad}}~[\mathrm{L}_{\odot}]$   &   log-uniform &   $[10^5,10^{12}]$\\
    $L_{\mathrm{H}\alpha,\mathrm{narrow}}~[\mathrm{L}_{\odot}]$ &   log-uniform &   $[10^5,10^{12}]$\\
    $lsf\_scale$                                                &   uniform     &   $[1,2]$\\
    $noise\_scale$                                              &   log-uniform &   $[0.1,10]$\\\hline\hline
    \end{tabular}
    \end{center}
    \footnotesize
    \par\noindent -- \texttt{Note:} The spectra are forward-modeled in rest-frame to reduce uncertainties.
\end{table*}

\end{appendix}

\end{document}